# Performance Evaluation of Incremental K-means Clustering Algorithm

Sanjay Chakraborty , N.K. Nagwani
National Institute of Technology (NIT) Raipur, CG, India

**ABSTRACT**

The incremental K-means clustering algorithm has already been proposed and analysed in paper [*Chakraborty and Nagwani, 2011*]. It is a very innovative approach which is applicable in periodically incremental environment and dealing with a bulk of updates. In this paper the performance evaluation is done for this incremental K-means clustering algorithm using air pollution database. This paper also describes the comparison on the performance evaluations between existing K-means clustering and incremental K-means clustering using that particular database. It also evaluates that the particular point of change in the database upto which incremental K-means clustering performs much better than the existing K-means clustering. That particular point of change in the database is known as 'Threshold value' or '% delta ($\delta$) change in the database'. This paper also defines the basic methodology for the incremental K-means clustering algorithm.

## 1. INTRODUCTION

Incremental clustering is a very important and needful approach for today's busy life. It is such a concept by which incremental data can be handled efficiently in a database. Today most of the databases are dynamic in nature (such as WWW and data warehouses) means data are inserted into the database and deleted from the database frequently. To save lot of time, cost and effort a new incremental K-means clustering algorithm has been already proposed in paper [*Chakraborty and Nagwani, 2011*]. This paper mainly discusses the basic methodology of the incremental K-means clustering algorithm, describes some illustrative examples and finally analyses the experimental results of this approach. So, the main objective of this paper is to define and evaluates that particular (%$\delta$) change of the database up to which incremental K-means clustering performs much better than the existing K-means clustering.

The rest of this paper is organized as follows. Section 2 discusses methodology of incremental clustering technique. The proposed model and Illustrative examples are reported in section 2.1 & 2.2 respectively. Section 3 describes the experimental results of the incremental K-means clustering. Section 3.1 and 3.2 describes the experimental setup and performance evaluations respectively. Section 4 concludes with a summary of those clustering techniques. Section 5 describes the references.

## 2. METHODOLOGY

Methodology of this approach consists of the combination of the proposed model and its illustrative examples. The proposed model simply describes the pictorial view of the approach and the illustrative examples discuss its logical and mathematical concepts.

### 2.1. Proposed Model

Problem: In dynamic databases, if new transactions (records/rows) are appended as time advances. It is an incremental algorithm, used to deal with this problem. The proposed algorithm identifies the value of percentage of size of original database x, which can be added to original database. Now there might be two following cases:

1. Up to x% change in the original database, better to use previous result.

2. For more than x %, rerun the algorithm again.
Solution: As a result, Incremented K-means will provide faster execution than the algorithms used previously K-means because the number of scans for the database will be decreased.

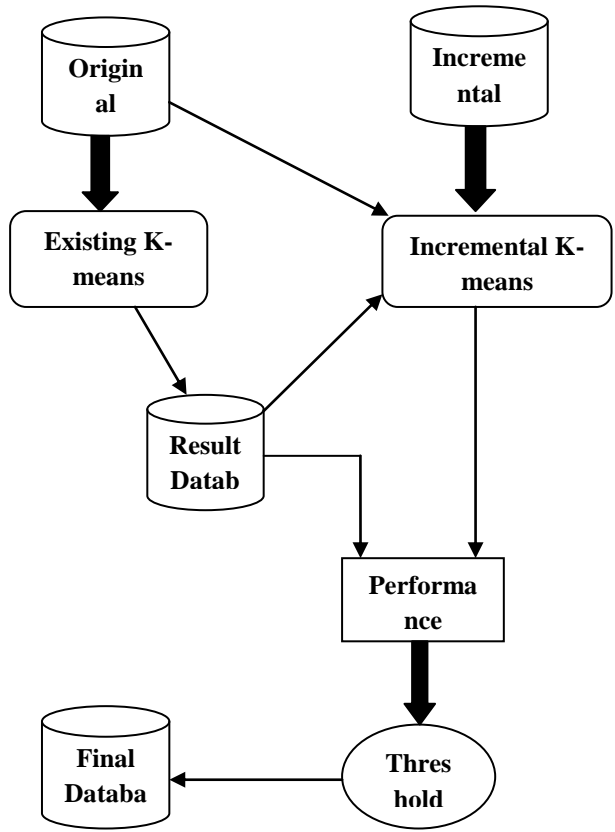

**Figure 1:** Performance evaluations of incremental k-means clustering

"Figure 1" describes the main aim of this paper clearly. First, the existing K-means clustering algorithm (developed in Java) is applied on the original air-pollution database and the result is stored in a result database using mysql. Now, the new data is coming and inserted into that existing database which in turn to be known as incremental database. The incremental K-means clustering algorithm is applied on that incremental data after collecting necessary information from the result database. Thus the new data is directly inserted into the existing database without running the K-means algorithm again and again. Finally the results of these two are compared and also evaluate the performance as well as its correct threshold value. This concept is more clearly discussed by the mathematical examples given below.

### 2.2. Illustrative Examples

The proposed model can be explained with the help of following two examples:

**Example.1**. Suppose there is a set of data objects, such as A(15),B(7),C(8),D(11),E(5),F(14),G(3),H(1). Assume that the points A, E and H are three cluster centers. Form clusters properly using k-means algorithm. Suppose two new data I(17) and S(9) are inserted and also two data items D(11) and F(14) are deleted later. Then show how this algorithm will behave?

**Sol**: Suppose the initial cluster centers are A(15), E(5), H(1). Computation is done below using Manhattan distance metric.($D=|x_1-x_2|+|x_2-x_3|+.......$).

First iteration:

| Data Items | A(15) | E(5) | H(1) | New Clusters |
|---|---|---|---|---|
| A(15) | 0(min) | 10 | 14 | 1 |
| B(7) | 8 | 2 | 6 | 2 |
| C(8) | 7 | 3 | 7 | 2 |
| D(11) | 4 | 6 | 10 | 1 |
| E(5) | 10 | 0 | 4 | 2 |
| F(14) | 1 | 9 | 13 | 1 |
| G(3) | 12 | 2 | 2 | 3 |
| H(1) | 14 | 4 | 0 | 3 |

Now,                  #items     Mean
Cluster1={A(15),D(11),F(14)}=3     = 13.3
Cluster2={B(7),C(8),E(5)}     =3     = 6.7
Cluster3={G(3),H(1)}           =2     = 2

Based on these three means the items distances and their group of clusters can be computed. If there is no change in the $2^{nd}$ iteration of the data, then this algorithm is terminated. This algorithm is also known as 3-means clustering. According to the mean value, the new group of cluster items are:

**Second iteration:**

| Data Items | New Clusters |
|---|---|
| A(15) | 1 |
| B(7) | 2 |
| C(8) | 2 |
| D(11) | 1 |
| E(5) | 2 |
| F(14) | 1 |
| G(3) | 3 |
| H(1) | 3 |

⇨ No Change

Insertion:

Now, Two new data items I(17) and S(9) are inserted, then in the first approach those data are clustered directly after comparing with the means of existing clusters using Manhattan distance metric. Such as

I(17) => 3.7(min)   6.7   2   => cluster1
S(9) => 4.3   2.3   7   => cluster2

But in the second approach, if the k-means algorithm is run again from beginning with those two new data items then it results as follows:

| Data Items | A(15) | E(5) | H(1) | New Clusters |
|---|---|---|---|---|
| A(15) | 0(min) | 10 | 14 | 1 |
| B(7) | 8 | 2 | 6 | 2 |
| C(8) | 7 | 3 | 7 | 2 |
| D(11) | 4 | 6 | 10 | 1 |
| E(5) | 10 | 0 | 4 | 2 |
| F(14) | 1 | 9 | 13 | 1 |
| G(3) | 12 | 2 | 2 | 3 |
| H(1) | 14 | 4 | 0 | 3 |
| I(17) | 2 | 13 | 16 | 1 |
| S(9) | 6 | 4 | 8 | 2 |

⇨ The Result is same but the 2$^{nd}$ approach is more time-consuming and more effort-able compare to the first approach.

Deletion:
Now if two items D(11) and F(14) are deleted from the existing database then in the first approach after calculating from clusters new means(after deletion),

Now,                                #items
Cluster1= {A(15)}                    = 1
Cluster2= {B(7),C(8),E(5)}           = 3
Cluster3= {G(3),H(1)}                = 2

| Data Items | A(15) | E(5) | H(1) | New Clusters |
|---|---|---|---|---|
| A(15) | 0(min) | 10 | 14 | 1 |
| B(7) | 8 | 2 | 6 | 2 |
| C(8) | 7 | 3 | 7 | 2 |
| E(5) | 10 | 0 | 4 | 2 |
| G(3) | 12 | 2 | 2 | 3 |
| H(1) | 14 | 4 | 0 | 3 |

Now,                        #items         Mean
Cluster1= {A(15)}            = 1           = 15
Cluster2={B(7),C(8),E(5)}=3                = 6.7
Cluster3= {G(3),H(1)}        = 2           = 2

No change in the cluster.

⇨ Here also the 2$^{nd}$ approach is time consuming and more effort-able compare to the first one.

**Example.2**: Sometimes mean of a cluster depends on the dimensions of its using database. Suppose, a multidimensional database has four attributes so each cluster of that database must produce four centroids or means. Suppose, suppose an air pollution database has four attributes SPM (Suspended Particulate Matter), RPM (Respirable Particulate Matter), Nitrogen Dioxide ($NO_2$) and Sulphur Dioxide ($SO_2$). Assume the initial number of cluster is 3. And also assume that after the first iteration the means of each cluster are

|  | Cluster 0 | Cluster 1 | Cluster 2 |
|---|---|---|---|
| SPM | 24 | 32 | 15 |
| RPM | 22 | 42 | 20 |
| $NO_2$ | 12 | 32 | 9 |
| $SO_2$ | 14 | 27 | 12 |

Now if a new data is entered into the existing database with value SPM=21, RPM=8, NO2=9, SO2=12 then it first compares each of its attribute's distance with the attributes of existing cluster by the help of distance metric (Euclidean metric). And it will enter into that cluster where the distances are minimum. Such as,

$Cluster0 = \sqrt{(24-21)+(22-8)+(12-9)+(14-12)} = 4.7$

$Cluster1 = \sqrt{(32-21)+(42-8)+(32-9)+(27-12)} = 9.11$

$Cluster2 = \sqrt{(21-15)+(20-8)+(9-9)+(12-12)} = 4.2$ (minimum)

So the new data item should be entered into the 'Cluster 2' directly without rerunning the whole algorithm. Thus it saves our time and effort both.

## 3. EXPERIMETAL RESULTS

This paper implements the incremental K-means clustering approach and also computes its speed of processing over existing K-means clustering algorithm.

### 3.1. Experimental Setup

This experiment is done on air-pollution database with the help of Java language, Weka interface and other tools.

### 3.1.1 Air-pollution Database

This analysis is based on the observation of the air pollution data has been collected from "West Bengal Air Pollution Control Board" and the URL is- "http://www.wbpcb.gov.in/html/airqualitynxt.php". This database consists of four air-pollution elements or attributes and they are Suspended particulate matter (SPM), Respirable particulate matter (RPM), Sulphur dioxide ($SO_2$) and Oxides of Nitrogen ($NO_x$). Air pollution data of each day are collected and stored that record in an .arff (Attribute resource file format) file format. Finally use this database directly with the proposed K-means clustering algorithm. The detail database format is shown in the "Table 1".

| Date | SPM | RPM | $SO_2$ | $NO_X$ |
|---|---|---|---|---|
| 1/1/2009 | 357 | 183 | 12 | 95 |
| 2/1/2009 | 511 | 289 | 14 | 125 |
| 3/1/2009 | 398 | 221 | 10 | 101 |
| 4/1/2009 | 358 | 191 | 11 | 97 |
| 5/1/2009 | 329 | 175 | 11 | 101 |
| ……… | ....... | ……. | …… | …….. |

**Table 1:** Original air-pollution Database

### 3.1.2 Java
Java stands for "just avail vital abstraction". This research paper result analysis is based on the platform of JAVA1.5 compiler.java higher version (JAVA 7.0) is not used because of Generics problem. In this paper incremental K-means algorithm is developed in Java.

### 3.1.3 Weka
Weka (Waikato Environment for Knowledge Analysis) is the other open source API's (Application Programming Interfaces) to support the other functionalities. Weka is used for performing some data mining related operations.

### 3.1.4 Eclipse
Eclipse is used as a development IDE (Integrated Development Environment) for java and library of other technologies are added as external jar (Java Archives) in the eclipse.

### 3.1.5 Applying databases or Mysql
As per paper observation, these are mainly used to construct databases on which k-means algorithms are applied. All the experiments are performed on a 2.26 GHz Core i3 processor computer with 4GB memory, running on Windows 7 home basic.

### 3.2. Performance Evaluations
In the performance evaluation, both techniques involve computation of centroids where these centroids will be used to cluster the data. In the actual K-means clustering, the algorithm is applied on the air-pollution dataset and form clusters based on the nearest distance of the data from predefined centroids. But in case of dynamic environment, when new data is entering into the air-pollution database, the incremental K-means clustering is applied. This technique performs its operations on the existing clusters and clustered the new data directly by using the nearest distance between the new data and the centroids of the existing clusters. Both of these techniques use Euclidean distance measure function in this experiment.

At first, initialize the total number of clusters are five, then the actual K-means clustering algorithm is running on four dimension attributes based air-pollution database. So, each cluster consists of four objects. The result are stored into two different databases, they are as follows.

| clusterid | clustSPMmean | clustRPMmean | clustSOmean | clustNOmean |
|---|---|---|---|---|
| cluster0 | 321.376238 | 164.366337 | 10.128713 | 92.415842 |
| cluster1 | 252.600000 | 118.562500 | 8.425000 | 72.187500 |
| cluster2 | 93.458824 | 36.176471 | 5.158824 | 41.523529 |
| cluster3 | 165.196721 | 75.983607 | 6.704918 | 57.04918 |
| cluster4 | 388.943182 | 202.022727 | 12.034091 | 107.102273 |

**Table 2:** Means of five clusters

| clusternumber | distancefunction | clusteriteration | squareError |
|---|---|---|---|
| 5 | Euclideandistance | 35.0000 | 12.53647 |

**Table 3:** Different parameters of the actual K-means clustering algorithm

Table 2 defines the means of the five clusters after applying existing K-means clustering algorithm on the air-pollution database. Here the required time is measured using currentTimeMillis() method of java. After measuring time for the change of data in the database, the following table 4 and figure 2 can be formed

| Original Data | Time (ms) |
|---|---|
| 1000 | 156ms |
| 1100 | 172ms |
| 1200 | 172ms |
| 1300 | 187ms |
| 1400 | 188ms |
| 1500 | 188ms |
| 1600 | 203ms |
| ……. | ……. |

**Table 4:** Time vs. data in actual K-means clustering

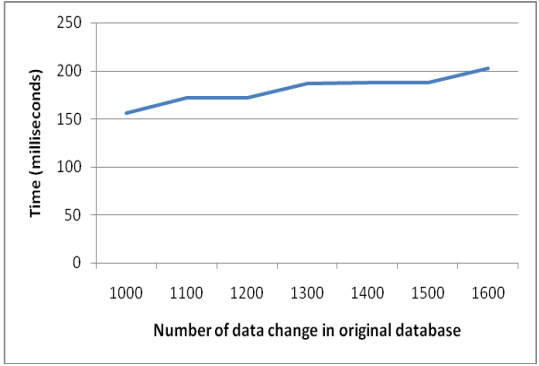

**Figure 2:** Graph for actual K-means result

Figure 2 describes that how the time slowly increases with the increases of data in the original database. Now when the new data are inserted into the old database, then for that new data the proposed incremental K-means clustering algorithm is applied. This algorithm directly clustered the new coming data without rerunning the K-means algorithm by comparing those data with the means of existing clusters (from table 2). The relation of the required time against that new incremented data is shown by the table 5 and the figure 3 below.

| Incremental Data | Time (ms) |
|---|---|
| 100 | 47 |
| 200 | 94 |
| 300 | 125 |
| 400 | 172 |
| 500 | 178 |
| 600 | 218 |
| …….. | …….. |

**Table 5:** Time vs. incremented data in incremental K-means clustering

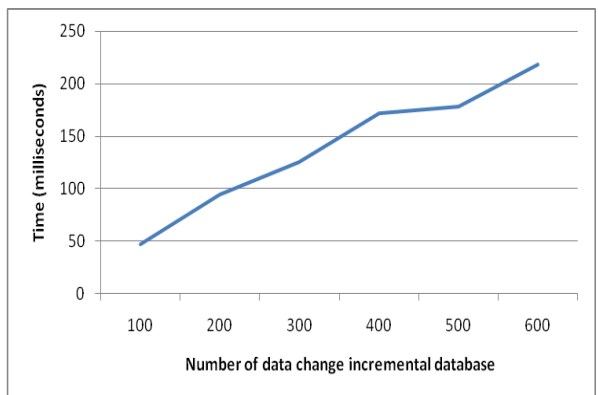

**Figure 3:** Graph for incremental K-means result

Figure 3 describes how the time rapidly increases with the increases of data in the incremental database.

Now, it can easily calculate after combining the above two results that for what % of delta ($\delta$) change in the database up to which the incremental K-means clustering behaves better than the actual K-means clustering. First calculate all the delta changes of this database by the help of following formula.

%$\delta$ change in DB = $\frac{(NEW\ DATA - OLD\ DATA)}{OLD\ DATA} \times 100$    [1]

| Actual Time(ms) | %$\delta$ change in the database | Incremental Time(ms) |
|---|---|---|
| 172 | $\delta_1 = \frac{(1100-1000)}{1000} \times 100 = 10\%$ | 47 |
| 172 | $\delta_2 = 20\%$ | 94 |
| 187 | $\delta_3 = 30\%$ | 125 |
| 188 | $\delta_4 = 40\%$ | 172 |
| 188 | $\delta_5 = 50\%$ | 178 |
| 203 | $\delta_6 = 60\%$ | 218 |
| …….. | ……… | ……. |

**Table 6:** Time vs. %$\delta$ change in DB for both actual and incremental K-means

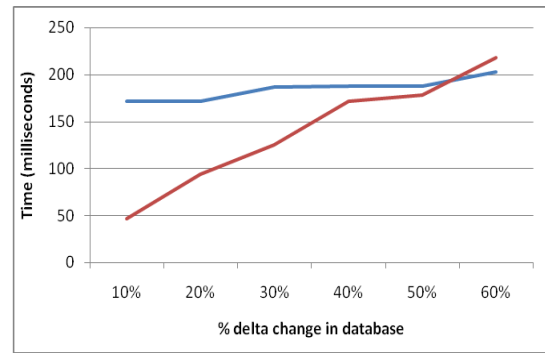

**Figure 4:** Graph for actual K-means vs. incremental K-means

From the above figure, it can be easily mentioned that the threshold value upto which the proposed K-means clustering behaves better than the existing one is 57% [Threshold value=57%]. But after that threshold value the actual clustering technique behaves better compare to the incremental clustering.

## 4. CONCLUSION

In this paper the performance evaluation of a proposed incremental K-means clustering algorithm is established. This performance measure and compare the performance with the existing K-means clustering is also presented in this paper clearly. The proposed technique is implemented using open source technology java, weka and air-pollution dataset is selected for the experiment.